\documentclass[english]{elsarticle}
\usepackage{float}
\usepackage{graphicx}
\usepackage{lineno,hyperref}
\usepackage{amsmath}
\usepackage{babel}
\usepackage[utf8]{inputenc}
\journal{Annals of physics}

\newcommand{\bra}[1]{\left\langle {#1} \right|}
\newcommand{\ket}[1]{\left|  #1 \right\rangle}

\begin{document}

\begin{frontmatter}

\title{Quantum enhanced precision in a collective measurement}

\author[mymainaddress,mycurrentaddress]{H. M. Bharath\corref{mycorrespondingauthor}}
\cortext[mycorrespondingauthor]{Corresponding author}

\ead{bharath.hm@gatech.edu}

\author[mymainaddress]{Saikat Ghosh}

\ead{gsaikat@iitk.ac.in}

\address[mymainaddress]{Department of Physics,
Indian Institute of Technology, Kanpur,
208016, India}
\address[mycurrentaddress]{School of Physics, Georgia Institute of Technology, Atlanta, Georgia-30332, (U.S.A)}

\begin{abstract}
We explore the role of \textit{collective measurements} on precision in estimation of a single parameter. Collective measurements are represented by observables which commute with all permutations of the probe particles. We show that with this constraint, quantum bits(qubits) outperform classical bits(non-superposable bits) in optimizing precision. Specifically, we prove that while precision in a collective measurement is loosely bounded by $O\left(\frac{1}{N}\right)$ for $N$ classical bits, using qubits it is tightly bounded by $O\left(\frac{1}{N^2}\right)$. This bound is consistent with quantum metrology protocols with the collective measurement requiring an entangled probe state to saturate. Finally, we construct a canonical measurement protocol that saturates this bound. 
\end{abstract}

\end{frontmatter}
 
\section{Introduction:} 
Experiments in precision metrology for estimation of a parameter are usually modelled in three distinct stages: 1) a choice of an initial state of probe particles, 2) preparation of a final probe state through evolution and 3) a suitably chosen measurement performed on this final state \cite{Giovan-06}. While the \textit{true} value of the parameter to be estimated, is imprinted on the probe state during the second stage, the third stage returns an estimate ($z$) that approximates the true value, ($x$), within a finite precision ($\delta$): $x \in [z-\delta,z+\delta]$. The final precision of estimation depends on each of the three stages. Accordingly, understanding and optimizing limits of precision for a fixed number of probe particles and under realistic experimental constraints, have been a major focus of research over past several decades \cite{Holevo_book, Helstrom_book, Wiseman_book, Hayashi_book}. 
   
Recent work has been focused on optimization of precision over choices of initial states, while keeping the second stage of state evolution fixed  \cite{Giovan-06, Giovan-11, Giovan-04, Toth-14}. In particular, for a Mach-Zender like interferometeric evolution, the second stage constitutes repeated application of a unitary operator. For $n$ such applications, it is known that the precision scales as $\frac{1}{n}$ in general (Heisenberg Limit) while the scaling is $\frac{1}{\sqrt{n}}$ when the initial state is restricted to be separable or classically correlated (standard quantum limit) \cite{Massar-2, Giovan-06, Giovan-12, Massar-3, Hayashi-1, Hayashi-2,  Wineland-92, Losses}. For few special cases of linear sequential evolution, $n$ is equivalent to the number of probe particles $N$, but not in general. For these strategies, the final measurement, in the form of differences of photo-currents at the output, is also kept fixed and it has been shown that if the input probe states are classically correlated, one cannot gain any further quantum advantage through entangling measurements \cite{Giovan-06}. Progress has also been made in modifying the second stage of state preparation using non-linear Hamiltonians \cite{NonLin-1, NonLin-2, NonLin-3}. With the evolution constrained to be interferometric, these approaches thereby provide a basis for comparing entangled and separable input probe states as resource for enhancing precision.

One can also set a crude lower bound on precision through an optimization over all possible initial states, final state preparation and measurements, keeping the number of probe particles constant. The corresponding bound on precision scales inversely with the number of \textit{distinguishable states} of the probe particle space and is independent of the nature of the dynamics, be it classical or quantum. Interestingly, this implies that in general, quantum bits or qubits as probe particles offer no particular advantage over classical (non-superposable) bits or cbits. The precision in this case is tightly bounded by $\delta \geq \frac{1}{2^N}$, for both classical and quantum bits \cite{Holevo}.

In this work,  we compare cbits versus qubits as probe particles, with no constraint on initial state or on preparation of the final state. However, we constrain the measurement to be a \textit{collective measurement} over all the probe particles. This constraint is equivalent to the measurement remaining invariant under particle permutation. We prove that for any such measurement using $N$ qubits, the optimal precision($\delta$) scales as 
\begin{equation}\label{quantum collective measurement}
\delta \geq \frac{1}{N^2}.
\end{equation}
We also demonstrate a canonical measurement protocol that saturates the above bound. 

For cbits as probes, the corresponding bound is $\delta \geq \frac{1}{N}$ (Table I). This bound is however not necessarily tight. For uncorrelated cbits, the central limit theorem sets a tighter bound, $\delta \geq \frac{1}{\sqrt{N}}$. Although we prove that $\delta \geq \frac{1}{N}$, a protocol that saturate this bound needs to be established. 

It can be noted that these results are consistent with general observations made in Ref. \cite{Giovan-06}: the quantum advantage reported here for collective measurements requires entangled probe states or qubits.  There is also practical motivation for understanding and implementing collective measurements on large ensembles of particles. For example, measurements on cold or ultra-cold atoms are becoming increasingly sophisticated and can therefore provide an alternative route for a significant quantum advantage in optimizing precision. 

\begin{table}[h]\label{Table 1}
\begin{tabular}{ c c c }
\hline
\textbf{Measurement}  & \textbf{Classical} & \textbf{Quantum} \\
\hline
Individual measurement & $ \frac{1}{2^N} $ & $\frac{1}{2^N}$ \\

Collective measurement with distinguishable particles  & $\frac{1}{N} $  & $\frac{1}{N^2}$ \\

Collective measurement with identical particles  & $\frac{1}{N} $  & $\frac{1}{N}$ \\
\hline
\end{tabular}
\caption{Lower bounds on precision($\delta$) for collective and individual measurements using $N$ qubits and cbits. As we show here, a quantum advantage appears only for a collective measurement}
\end{table}

To prove this result, we first show that in any measurement protocol, $\delta \geq \frac{1}{M}$, where $M$ is the total number of distinct possible outcomes. Though the bound is intuitive and has been used before \cite{Kitaev, Massar-3, Holevo_book}, we provide a short formal proof to keep the draft self contained. Next, we show that in protocols that use a collective measurement, $M\leq N^2$, thus proving \eqref{quantum collective measurement}. Finally, we analyse two well known but contrasting schemes of phase estimation from the perspective of counting distinguishable states, estimating $M$ for each case.

\begin{figure}
\includegraphics[scale=0.56]{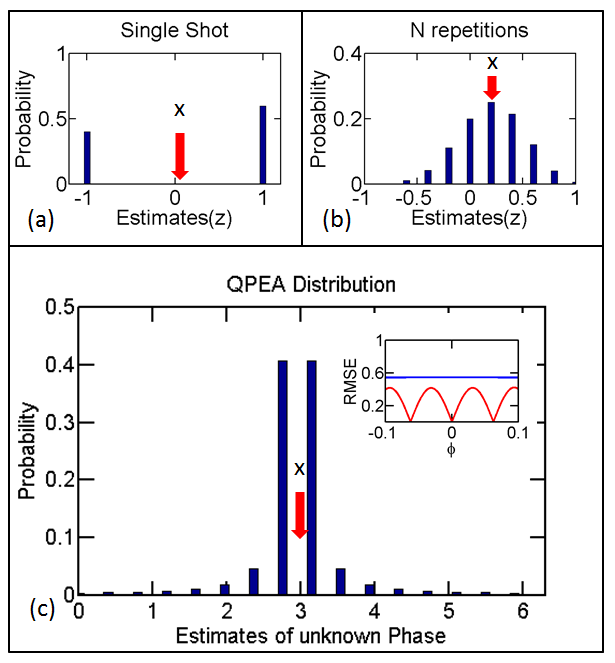}
\caption{ (a) The probability distribution in a single shot measurement. (b) The probability distribution for $N$ uncorrelated repetitions. (c)Probability distribution for QPEA. The RMSE is shown in the inset}
\end{figure}

\section{Bound on Precision: $\delta\geq 1/M$}

Any measurement model that uses \textit{limited resource} also predicts a finite set of probable values for the estimate: $z\in \{z_1, z_2 \cdots z_M\}$ with corresponding probabilities or likelihood functions $\{P_1(x),P_2(x) \cdots P_M(x)\}$, expressed as functions of the true value($x$). Here each value of $z$ corresponds to a probable final experimental outcome, where the experiment might itself include any number of repetitions of a basic measurement (Fig.1) \cite{Holevo_book, Delta-1}. We elucidate this further with a standard example.

\textit{Example:} 

An unknown phase with a true value $\phi$ in a Mach-Zender interferometer is inferred from a measurement of a parameter $ x= \cos(\phi)$  \cite{Giovan-04, Wiseman_book, Giovan-11}. For a single-photon incident on one of its input ports ($a,b$) with an initial state $\ket{\psi}=\ket{1}_a\ket{0}_b$, the output evolves to a state $\ket{\psi_{out}}_\phi = e^{i\phi/2}[cos(\phi/2)\ket{\psi}-i sin(\phi/2)\tilde{\ket{\psi}}]$, where, $\tilde{\ket{\psi}} = \ket{0}_a\ket{1}_b$. Note that the output state has a parametric dependence on a phase-element inserted in one of the interferometer arms. A choice of the measured observable($\hat{O}$) corresponds to the difference counts at the output ports: $\hat{O}=\ket{\psi}\bra{\psi}-\tilde{\ket{\psi}}\tilde{\bra{\psi}}$. It has an expected value: $\bra{\psi_{out}} \hat{O}\ket{\psi_{out}} = \cos(\phi)$. 

In this \textit{single shot} experiment, the probable values of the estimate are two distinct numbers: $z\in \{-1,1\}$ i.e., $M = 2$, with associated probabilities $P_{-1}(\phi) = \sin^2(\phi/2)$ and $P_{+1}(\phi) = \cos^2(\phi/2)$. While the numbers are the eigenvalues of $\hat{O}$, the probabilities reflect a parametric dependence on $\phi$ (Fig.1a).

\textit{N repetitions} with $N$ uncorrelated photons, increase the number of probable values ($M$) of the estimate. The measured operator $\hat{\mathcal{O}}$ is now described in a $2^N$ dimensional Hilbert space as: $\hat{\mathcal{O}}= \frac{1}{N}[(\hat{O}\otimes 1 \otimes \cdots \otimes 1)+ \cdots + (1 \otimes \cdots \otimes \hat{O})]$. It has $M = N+1$ \textit{distinct} eigenvalues:  $z\in \{-1,-1+\frac{2}{N}\cdots -1+\frac{2k}{N}  \cdots 1\} $ with corresponding probabilities corresponding to binomial distribution (Fig.1b). However, one can note that the expected value still remains $\langle\hat{\mathcal{O}}\rangle = \cos(\phi)$. 

Following the usual approach, for an experiment that returns a specific outcome $z_r$ (after $N$-repetitions), one infers the true value from the estimate by inverting the relation $|\cos(\phi)-z_r| \leq \sigma(\phi)$ i.e.,
\begin{equation*}
\cos(\phi) = \frac{N z_r}{N+1}\pm \sqrt{\frac{1}{N+1}-\frac{Nz_r^2}{(N+1)^2}},
\end{equation*}
which for large $N$, reduces to $\cos(\phi) \in [z_r \pm \frac{\sqrt{1-z_r^2}}{\sqrt{N}}]$. It can be noted that the precision, in this case, $\delta_r = \frac{\sqrt{1-z_r^2}}{\sqrt{N}} $ depends on the corresponding estimate $z_r$. However, in the allowed range of $x\in[-1,1]$, the maximum it can take sets the overall precision for the model. Equivalently, $\delta = \delta_{max} \sim \frac{1}{\sqrt{N}}$.  

\textit{A bound on precision:} Without loss of generality, here and for all subsequent cases, the range of $x$ is scaled to $\pm 1$. It is therefore unit-less. 

One can intuitively appreciate that if a model predicts $M$ probable outcomes, the precision should be bounded below by $\frac{1}{M}$. In the above example, the overall precision for a model was set by the worst it can do over the possible range of $x$; i.e., by taking the maximum of all $\delta_r$. Furthermore, $\delta_r$s together cover the probable range of $x$, equivalently, $\sum^M \delta_r \geq 1$. Consequently, $\delta$ always satisfies $\delta \geq \frac{1}{M}\sum_{r=1}^M \delta_r \geq \frac{1}{M}$ which sets a lower bound for precision: 
\begin{equation}\label{bound on delta}
\delta \geq \frac{1}{M}
\end{equation} 
This bound is saturated by a trivial distribution, with $P_r(x) = 1$ for the estimate $z_r$ that is \textit{closest} to the natural value $x$ and zero for all others. Such a distribution corresponds to a measurement that always returns the same estimate $z_r$. The corresponding inferred range for the true value is $x \in [\frac{z_{r-1}+z_{r}}{2},\frac{z_{r}+z_{r+1}}{2}]$. Since $x$ is unknown, it is straightforward to note that precision is optimized when all $z_i$s are equispaced, saturating the bound $\delta= 1/M$. Clearly, this distribution represents a biased estimator \cite{Delta-1}.  This completes the proof. As we show below, this bound is saturated for Quantum Phase Estimation Algorithm (QPEA)(Fig. 1c)  \cite{Kitaev}.

Connections between measurement precision and information theory identifying mutual information as a measure of precision have long been explored  \cite{Preskill-1, Massar-1, Brauns-92}. It is straightforward to note that the mutual information($I$) \cite{Shannon, Nielsen} between $x$ and $z$ is bounded as $I\leq \log(M)$, with the bound saturated again for the above biased distribution. This implies that from the estimate, the true value can be estimated up to $\log(M)$ bits or equivalently, to a precision of $1/M$. Accordingly, here we propose  $e^{-I}$ as an alternate measure of precision for any general distribution, the measure being independent of RMSE or any other statistical ``width" functions(table -II).

It follows that the precision in a measurement protocol that uses $N$ cbits is bounded by $\delta \geq \frac{1}{N+1}$. While we have shown that there exists a probability distribution, there may not be a complete protocol using cbits, that saturates the bound. In contrast, for qubits, we show that a measurement protocol can always be constructed that saturates the above bound of $1/M$.

\subsection{Canonical measurement protocol:}(see \cite{Kitaev, Massar-3, Holevo_book} for more details)
Quantum mechanically, the measured quantity is represented by an observable (the generality follows from Neumark's theorem \cite{Neumark}, that every POVM can be dilated into a PVM in a larger Hilbert space) and therefore, $M$ corresponds to the number of \textit{distinct eigenvalues} of the observable. Here we provide a canonical measurement protocol for every observable $\hat{O}$, that saturates the above bound on precision.

Let us denote the eigenvalues and the eigenstates of $\hat{O}$ by $\lambda_1, \lambda_2 \cdots \lambda_M $ and $|\lambda_1\rangle, |\lambda_2\rangle \cdots |\lambda_M\rangle$ respectively. 
One can then construct a set of orthonormal states $\{|\tilde{\lambda}_1\rangle, |\tilde{\lambda}_2\rangle \cdots |\tilde{\lambda}_M\rangle\}$ by applying a quantum Fourier transformation \cite{Nielsen} to the eigenstates of $\hat{O}$:
\begin{equation}
|\tilde{\lambda}_k\rangle= \sum_{j=1}^M e^{i\frac{2\pi j k}{M}}|\lambda_j\rangle
\end{equation}

To construct a protocol for the estimation of an unknown phase $\phi$, let us choose the initial state, $|\psi_i\rangle = |\lambda_1\rangle$ and the Hamiltonian:
\begin{equation}
H= \sum_{i=k}^M k|\tilde{\lambda}_k\rangle \langle \tilde{\lambda}_k|
\end{equation}

The final state is then given by $|\psi_f\rangle = e^{iH\phi}|\psi_i\rangle $. After performing a measurement of the observable $\hat{O}$ on the final state,  we obtain an eigenvalue $\lambda_k$ as the outcome. Choosing the corresponding estimate as $\hat{\phi}= 2\pi \frac{k}{M}$, it is straightforward to check that the corresponding precision is $\frac{2\pi}{M}$ which is equivalent to a relative precision of $\frac{1}{M}$. The resource count\cite{Giovan-11} for this protocol is $n = M$ and therefore, this protocol also saturates the Heisenberg limit.  

We next use this bound on precision to prove our main result.

\section{Collective Measurements:}  For optimal precision, one needs to choose an observable with the largest possible number of distinct eigenvalues. In absence of any additional constraints on the model, the dimensionality of the Hilbert space $dim\{\mathcal{H}\}$ sets a trivial bound on $M$:
\begin{equation}\label{bound on M}
M \leq dim\{\mathcal{H}\} 
\end{equation}

Typically, $dim\{\mathcal{H}\}$ is exponential in the number of probe particles ($N$) used, suggesting a possible exponential scaling of the corresponding precision with $N$ \cite{NonLin-3}. For instance, a measurement using $N$ qubits can be precise upto $\frac{1}{2^N}$, as is the case with $N$ cbits. Measurements that saturate this bound necessarily employ \textit{individual measurements} on all of the probe particles. On the contrary, a \textit{collective measurement}, characterized by its invariance under particle permutations, is experimentally more accessible. For example, collective measurements in the form of Dicke super-radiant intensity of $N$ two level atoms or measurement of total spin of a cold atomic cloud do not need to measure each individual particles in the ensemble. 

A collective measurement using $N$ cbits can have at most $N+1$ distinct outcomes, restricting the precision to $\frac{1}{N}$. For $N$ qubits, a collective measurement is represented by an observable that remains invariant under all particle permutations. Therefore it commutes with all the unitary operators that represent particle permutations. The total spin operator $J_N^2 + J_N^{(z)}$ is one such observable representing a collective measurement and it typically has $\sim N^2$ distinct eigenvalues, allowing a precision up to $\frac{1}{N^2}$ with a clear quantum advantage. It may be noted that while the operator $J_N^2 + J_N^{(z)}$ is symmetric under particle exchange, its eigenstates are not symmetrized. Instead, each of its degenerate \textit{eigenspace} is invariant under particle exchange. Therefore, even though the measurement is collective, the particles still need to be distinguishable, in order to reach a precision of $\frac{1}{N^2}$. 

The question then is, what is the maximum number of distinct eigenvalues a collective measurement can have? In theorem-1 we prove that there is no other observable that represents a collective measurement, with a larger number of distinct eigenvalues than $J_N^2 + J_N^{(z)}$. We thus establish that a collective measurement using qubits can be precise only up to $\delta =\frac{1}{N^2}$.

To complete the picture, we note that for $N$ \textit{identical} qubits, the Hilbert space dimension is reduced to $N+1$ due to symmetrization. Thus, the precision is bounded by $1/N$ for identical particles (Table I).

\textbf{Theorem 1:} For a collective measurement using $N$ two level probe particles, the count of outcomes($M$) is tightly bounded by the following inequalities
 \begin{equation}
 \begin{split} \label{bound for collective measurements}
 M\leq \frac{(N+2)^2}{4} \qquad
 & \text{for even N}\\
 M\leq \frac{(N+1)(N+3)}{4} \qquad
 & \text{for odd N}.
\end{split}  
\end{equation}

The inequality is saturated for the observable, $J_N^2+\epsilon J_N^{(z)}$, where $J^2_N$ is the total spin operator for $N$ particles and $J_N^{(z)}$ is its $z$ component. $\epsilon$ is a number, small compared to the eigenvalues of $J_N^2$, such that simultaneous eigenspaces of $J_N^2$ and $J_N^{(z)}$ have distinct eigenvalues. 

The symmetric group $S_N$ (permutation group of $N$ particles) has a unitary representation in the Hilbert of these $N$ particles. An observable represents a collective measurement iff each of its eigenspaces are invariant under each unitary in the representation of $S_N$.  $M$ is therefore bounded above by the number of irreducible representations of $S_N$ in the $N$ particle Hilbert space(counting isomorphisms), in an irreducible decomposition. This number is independent of the decomposition itself \cite{representation_theory-1}. Therefore, the above theorem may be restated as:

\textbf{Theorem 1 (restated)}: The common eigenspaces of $J_N^2$ and $J_N^{(z)}$ are irreducible under $S_N$.

\textbf{Proof}: We denote the common eigenspace of $J_N^2$ and $J_N^{(z)}$ by $\mathcal{V}_{Njm}$, with eigenvalues $j(j+1)$ and $m$ respectively. Since $J_N^2$ and $J_N^{(z)}$ are invariant under $S_N$, these spaces too are invariant under $S_N$. We now prove that they are also \textit{irreducible} under $S_N$.

This theorem is proved by induction on $N$. It is easily verified to be true for $N=2$, i.e.,  $\mathcal{V}_{2jm}$ are irreducible under $S_2$. Assuming that it is true for $2,3 \cdots N-1 $, we prove that it is true for $N$, by showing that $\mathcal{V}_{Njm}$ has no proper subspace invariant under $S_N$. To begin, we decompose $\mathcal{V}_{Njm}$ in to subspaces invariant under $S_{N-1}$(the subgroup of $S_N$ consisting of permutations of the first $N-1$ particles.)
The operator $J^2_{N-1}\otimes I$ is invariant under $S_{N-1}$. Therefore, its two eigen spaces within $\mathcal{V}_{Njm}$ are invariant under $S_{N-1}$. We refer to them by: $\mathcal{U}_{j\pm\frac{1}{2}}$. They have a natural basis:
\begin{equation}
\begin{split}
\mathcal{U}_{j \pm \frac{1}{2}}:  & C_{\pm1}|j\pm \frac{1}{2}, m-\frac{1}{2}\rangle\otimes |\frac{1}{2},\frac{1}{2}\rangle \\
&+ C_{\pm 2}|j\pm \frac{1}{2}, m+\frac{1}{2}\rangle\otimes |\frac{1}{2},\frac{-1}{2}\rangle 
\end{split}
\end{equation}
here, $C_{\pm i}$ are Clebsch-Gordon coefficients given by,
$C_{\pm 1}=C_{m-\frac{1}{2},\frac{1}{2};m}^{j\pm \frac{1}{2},\frac{1}{2};j}$
$C_{\pm 2}=C_{m+\frac{1}{2},\frac{-1}{2};m}^{j \pm \frac{1}{2},\frac{1}{2};j}$.
These two spaces are not only invariant, but they are indeed irreducible under $S_{N-1}$. We use the induction hypothesis to prove this.
\begin{table}

\begin{tabular}{ l|l l l l l  }

\hline
\textbf{Measurement} & \textbf{RMSE}& \textbf{Precision($\delta$)}&\textbf{Mut. Info.(I) } & \textbf{$e^{-I}$} &  \textbf{\# of probes(N)}\\
\hline
SQL &  $\frac{1}{\sqrt{M}}$&$\frac{1}{\sqrt{M}}$&$\frac{1}{2}\log(M/(2\pi e))$ &$\frac{\sqrt{2\pi e}}{\sqrt{M}}$ & $M=N+1$\\
QPEA \cite{Kitaev}   & $\frac{\sqrt{8\log(2)}}{\sqrt{M}}$&  $\frac{1}{M}$ &$\log(M)-2(1-\gamma)$ &$\frac{e^{2(1-\gamma)}}{M}$ & $M=2^N$\\
Q-Metrology \cite{Giovan-06}  & $\sim \frac{1}{10\sqrt{ \nu}}$ & $\frac{1}{\sqrt{M}}$&$\sim \frac{1}{2}\log(M)$ &$\sim \frac{1}{\sqrt{M}}$ & $M=N^{\log(2\nu+1)}$\\
\hline
\end{tabular}
\caption{A comparison of precision and RMSE for a few standard cases (See appendix for details). Usual symbols for standard constants are used, $\gamma$ being the Euler --Mascheroni constant. The last column lists $M$ for each case in terms of the number of probe particles ($N$). For quantum metrology (Q-Metrology), $\nu$ is the number of classical repetitions as discussed in the text. }
\end{table}

\textbf{Lemma 1}: The spaces $\mathcal{U}_{j \pm \frac{1}{2}}$ are irreducible under $S_{N-1}$

\textbf{Proof:} Consider the map $\left(J_{N-1}^-\otimes \sigma_+ \right):\mathcal{U}_{j\pm \frac{1}{2}} \rightarrow  \mathcal{V}_{N-1,j\pm \frac{1}{2},m-\frac{1}{2}}\otimes |\frac{1}{2},\frac{1}{2}\rangle$. This map is invariant under $S_{N-1}$. Let $W \subset \mathcal{U}_{j \pm \frac{1}{2}}$ be a subspace invariant under $S_{N-1}$. We have, $\left(J_{N-1}^-\otimes \sigma_+ \right) W \subset \mathcal{V}_{N-1,j\pm \frac{1}{2},m-\frac{1}{2}}\otimes |\frac{1}{2},\frac{1}{2}\rangle$ is also invariant under $S_{N-1}$. Noting that, from the induction hypothesis, $\mathcal{V}_{N-1,j\pm \frac{1}{2},m-\frac{1}{2}}$ is irreducible under $S_{N-1}$ and $Dim(\mathcal{V}_{N-1,j\pm \frac{1}{2},m-\frac{1}{2}})=Dim(\mathcal{U}_{j \pm \frac{1}{2}})$, we conclude that $W=\mathcal{U}_{j \pm \frac{1}{2}}$ or $W=0$, and therefore, $\mathcal{U}_{j \pm \frac{1}{2}}$ are irreducible under $S_{N-1}$.

Returning to the main proof, let $W$ be a nonzero invariant subspace of $\mathcal{V}_{Njm}$. We prove that $W=\mathcal{V}_{Njm}$. Let us define the projection maps $\pi$, $\pi_{\pm}$ into $W$ and $\mathcal{U}_{j\pm\frac{1}{2}}$ respectively:

\textbf{Def:} $\pi: \mathcal{V}_{Njm}\rightarrow W$ and $\pi_{\pm}:  \mathcal{V}_{Njm} \rightarrow \mathcal{U}_{j\pm \frac{1}{2}}$

$\pi$ is invariant under $S_N$, and $\pi_{\pm}$ are invariant under $S_{N-1}$. Therefore, the map $\pi_{-}\pi\pi_{+}: \mathcal{U}_{j+\frac{1}{2}}\rightarrow \mathcal{U}_{j-\frac{1}{2}}$ is also invariant under $S_{N-1}$. By Schur's lemma, either $\pi_{-}\pi\pi_{+}=0$ or $\pi_{-}\pi\pi_{+}=\lambda I$ on $\mathcal{U}_{j+\frac{1}{2}}$. 

In the former case, $\pi_{-}\pi\pi_{+}=0$, it is straightforward to show that $W=\mathcal{U}_{j\pm \frac{1}{2}}$ or $W=\mathcal{V}_{Njm}$. But, $\mathcal{U}_{j\pm \frac{1}{2}}$ are not invariant under $S_N$, therefore,  $W=\mathcal{V}_{Njm}$.

In the latter case, $\pi_{-}\pi\pi_{+}=\lambda I$ means that there exists an isomormphism between the spaces $\mathcal{V}_{N-1,j\pm \frac{1}{2},m}$, that is invariant under $S_{N-1}$. However, this is impossible because these spaces have different symmetry properties. 

For instance, consider an interchange of the $(N-1)^{th}$ and $(N-2)^{th}$ particle. Under this permutation, the asymmetric subspaces of $\mathcal{V}_{N-1, j\pm \frac{1}{2},m}$ are respectively isomorphic to $\mathcal{V}_{N-3, j\pm \frac{1}{2},m}$. Therefore, if there exists an isomormphism between the spaces $\mathcal{V}_{N-1,j\pm \frac{1}{2},m}$, that is invariant under $S_{N-1}$, it means not only that the spaces $\mathcal{V}_{N-1, j\pm \frac{1}{2},m}$ are isomorphic to each other, but also, that $\mathcal{V}_{N-3, j\pm \frac{1}{2},m}$ are isomorphic to each other. However, this is easily seen to be impossible by considering their dimensions. The dimension of the space $\mathcal{V}_{N, j,m}$ is a Catalan number and can be expressed using binomial coefficients. The dimensions of $\mathcal{V}_{N-1, j\pm \frac{1}{2},m}$ are equal iff, $N=(2j+1)^2-1$. Therefore it is clearly impossible for both the pairs $\mathcal{V}_{N-1, j\pm \frac{1}{2},m}$ and $\mathcal{V}_{N-3, j\pm \frac{1}{2},m}$ to be simultaneously isomorphic to each other.

We conclude with an analysis of two well known examples, illustrating estimation of the quantity $M$ and application of the corresponding bound derived here. We start with QPEA. 

\textit{Illustrations:} QPEA \cite{Kitaev} is a classic example that saturates both \eqref{bound on M} and \eqref{bound on delta} employing individual measurement on each of the probe particles. An important building block for quantum computers \cite{Nielsen}, it has also been used in interferometric measurements for estimating an unknown phase ($\phi$) \cite{Exp-Kitaev}. 

QPEA uses $N$ qubits, while the measured observable can be identified as \cite{Kitaev, Nielsen}: $\mathcal{O}_{QPEA}=\sum_{k=0}^{2^N-1}\frac{2\pi k}{2^N}|k\rangle \langle k|$, with $\{|k\rangle\}$ representing the \textit{computational basis}(Fig 1.d). For this operator, $M = 2^N$ and, $\delta =\frac{1}{2^N}$. On the contrary, the RMSE is of the order $\sim \frac{1}{\sqrt{2^N}}$(inset of Fig.1c, Table II), therefore failing to capture the error estimate for QPEA \cite{Massar-3}.

\textit{Quantum Metrology  \cite{Giovan-06}} models estimate an unknown phase($\phi$) to a precision of $1/N$, saturating the Heisenberg limit. These models use a maximally entangled input state 
$\ket{\Psi_{in}}=\frac{1}{\sqrt{2}}[\ket{\psi}^{\otimes N}+\tilde{\ket{\psi}}^{\otimes N}]$ (here, $\ket{\psi}$ and $\tilde{\ket{\psi}}$ are as defined in example 1). The corresponding measured observable is: $\mathcal{O}'=\ket{\psi}\bra{\psi}^{\otimes N}-\tilde{\ket{\psi}}\tilde{\bra{\psi}}^{\otimes N}$ with an expected value: $\langle\mathcal{O}'\rangle= \cos(N\phi)$, thereby estimating $N\phi$ instead of $\phi$.  
This observable has three eigenvalues: $\{-1,0,+1\}$ which are the probable values of the estimate, with $M = 3$. Repeating this experiment $\nu$ times yields a precision of $\frac{1}{\sqrt{\nu}}$ for $N\phi$ while $\phi$ is determined only upto one significant digit \cite{UnknownPhase-1}. 

However, as pointed out by \cite{Giovan-06}, with a choice of $N=10^j/2\pi$, one obtains the $j^{th}$ decimal place for $\phi$ with a precision of $1/\sqrt{\nu}$. The estimation will be exact for choice of $\nu \geq 100$. Therefore, a precision $1/N$ can be recovered by combining a series of runs $j=1,2,3\cdots \log(N)$. The model predicts $M=N^{\log(2\nu +1)}$ probable values of the estimates or outcomes, which reduces to $M \sim N^2$ for $\nu=100$. 

It can be noted that in this model, the total number of $N\nu$ particles is divided into $N$ batches of $\nu $ particles each and the measurement accesses each batch individually, but is collective within the batch.

\textit{Conclusion:}  
According to Holevo's theorem \cite{Holevo}, which sets an upper bound on the mutual information between the true value and the estimate \cite{MutInfo_1, MutInfo_2}, an unconstrained measurement using $N$ quantum bits can be no more precise than the corresponding unconstrained measurement using classical bits. Here we have shown that under the constraint that the measurement is collective, one do obtain a genuine quantum advantage. Qubit states posses additional properties of superposition and entanglement. From a theoretical perspective, we have identified one of the manifestations of these properties in the context of metrology. The generalization of our result higher spin systems can reveal even stronger quantum enhancement in precision.  To be able to realize such an enhancement requires the ability to discriminate every pair of Dicke states and a probable route to do so would be to use their superradiance properties \cite{Dicke}.

\textbf{Acknowledgments}

We acknowledge J. Kolodynski  and V. Ravishankar for fruitful discussions.


\bibliographystyle{unsrt}
\bibliography{Bibliography}

\section{Appendix: Entries of Table 2}

In this section we outline the calculations for the RMSE and Mutual Information(I) of Table-I in the main text. We start with a brief discussion on the definition of mutual information for our case:
 
\subsection*{Mutual Information}

Mutual information between $x$ and $z$ is defined in terms of the joint probability distribution $P(x,z_r)$. One can note that here while $x$ is a continuous variable, $z \in \{z_1 \cdots z_M\}$ is discrete. Accordingly, sum over $x$ is an integral in the range $x: [-1,1]$ and for $z$, it is a simple summation, from $r=1$ to $r=M$. The mutual information is then:  
\begin{equation}\label{def mut info}
\begin{split}
I=&\sum_{r}\int dx P(x,z_r)\log(P(x,z_r))\\
&-\sum_r\left(\int dx P(x,z_r)\right)\log\left(\int dx P(x,z_r)\right)-\int dx\left( \sum_r P(x, z_r)\right)\log\left( \sum_r P(x, z_r)\right) .
\end{split}
\end{equation}
For a measurement of an unknown parameter, the ``true value" $x$ is equally likely to be anywhere in $[-1,1]$. Therefore, the joint probability distribution is $\frac{1}{2}P_r(x)$, where $P_r(x)$ is the conditional probability $P(z_r|x)$. Substituting, one obtains a simpler form:

\begin{equation}\label{mut info}
I=\log(M)+\frac{1}{2}\sum_{r}\int dx P_r(x)\log(P_r(x))
\end{equation}
Here we have used
\begin{equation}
\begin{split}
P(x,z_r)=&\frac{1}{2}P_r(x)\\
\int dx P(x,z_r)= &1/M \\
\sum_r P(x,z_r)=&1/2
\end{split}
\end{equation}
These expressions follow again from the fact that $x$ is equally likely to be anywhere in the interval $[-1,1]$. We now take up each of the examples and calculate the mutual information and RMSE for each:

\subsection{SQL(Binomial)}
\subsubsection{Mutual Information}
For the binomial distribution considered in the very first example, we have, $M=N+1$ and $P_r(x)=\left( \begin{array}{c}N \\ r\end{array} \right)(\frac{1+x}{2})^{(N-r)}(\frac{1-x}{2})^r$. Using Stirling's approximation, we have:
\begin{equation}
\log(n!)= n\log(n)-n+\frac{1}{2}\log(2\pi n) + \epsilon(n)
\end{equation}
The remainder $\epsilon(n) \sim \frac{1}{12 n}$ can be ignored, since it is found to contribute nothing to the leading order. With this approximation, we have:
\begin{equation}\label{bino}
\begin{split}
\log(P_r(x))=&N\log(N)-N\log(2)+\frac{1}{2}\log(N)-\frac{1}{2}\log(2\pi)\\
&-\frac{1}{2}\log(r)-\frac{1}{2}\log(N-r)-r\log(r)-(N-r)\log(N-r)\\
&+(N-r)\log(1-x)+r\log(1+x)
\end{split}
\end{equation}
Evaluating the integral and sum for the constant terms in the above equation (the first line on the RHS), we obtain:
\begin{equation}\label{first line}
 2N\log(N)-2N\log(2)+\log(N)-\log(2\pi).
\end{equation}
After performing the integral over $x$, we are left with a sum over $r$ for the terms that depend only on $r$ (second line in the RHS):
\begin{equation}
\frac{2}{N+1}\sum_r (1+2r)\log(r)
\end{equation}
This can be evaluated by approximating with an integral. Accordingly, we obtain:
\begin{equation}\label{second line}
\sum_r(2r+1)\log(r)= 2N\log(N)+2\log(N)-N+1
\end{equation}
The last line of \eqref{bino} consists of terms that depend both on $r$ and $x$. The two terms are identical after the sum and integral, by symmetry. Observing that $\sum_r r P_r(x)= N\frac{1-x}{2}$, we obtain
\begin{equation}\label{third line}
N\int (1+x)\log(1+x)dx= 2N\log(2)-N
\end{equation}
Substituting \eqref{first line}, \eqref{second line} and \eqref{third line} in \eqref{mut info} we obtain, 
\begin{equation}
\boxed{
I= \log(N)+\frac{1}{2}(-\log(N)-\log(2\pi e))= \frac{1}{2}\log(N)-\log(\sqrt{2\pi e})}
\end{equation}
\subsubsection{RMSE}

By law of large numbers, the RMSE is,
\begin{equation}\boxed{
\sigma = \sqrt{\frac{1-x^2}{N}}
}
\end{equation}

\subsection{Quantum Phase estimation}
\subsubsection{Mutual Information}

For this case, $M=2^N$, where $N$ is the number of particles. The probable outcomes are $\{\frac{2\pi r}{M} |r =0,1\cdots M-1\}$. The probabilities in terms of the unknown phase $\phi$ is given by 
\begin{equation}\label{prob kit}
P_r(\phi)= \frac{1}{M^2}\left|\frac{1-t_r^M}{1-t_r}\right|^2
\end{equation}

with $t_r$ are complex numbers defined as: $t_r= e^{i(\phi-\frac{2\pi r}{M})}$. Note that $\phi$ belongs to a range of $[0,2\pi]$. Therefore, the factor of $\frac{1}{2}$ in \eqref{mut info} is replaced by $\frac{1}{2\pi}$. Taking logarithm of the probability, we obtain:
\begin{equation}\label{log prob kit}
\begin{split}
\log(P_r(\phi))=& -2\log(M) \\
& \log(1-t_r^M) +\log(1-t_r^{-M})\\
& -\log(1-t_r)-\log(1-t_r^{-1})
\end{split}
\end{equation}

The first line in \eqref{log prob kit} is a constant. Therefore, sum over r and integral over $\phi$ yields $-4\pi \log(M)$. The second line consists of terms independent of $r$, since, $t_r^M= e^{iM\phi}$. Therefore, we may sum over $r$ and use a taylor expansion to integrate over $\phi$, since $|t_r|=1$. We thus have, 
\begin{equation}
\int \log(1-e^{iM\phi})=  -\sum_{k=1}  \int\frac{e^{iMk\phi}}{k} d\phi=0
\end{equation}
Thus, the second line does not contribute anything. The third line in \eqref{log prob kit} consists of terms that depend on both $r$ and $\phi$. After a Taylor expansion, one can express:
\begin{equation}
\log(1-t_r)= -\sum_{k=1}^{\infty} \frac{t_r^k}{k} 
\end{equation}
The probability distribution \eqref{prob kit} can also be rewritten as:
\begin{equation}
P_r(\phi)= \frac{1}{M^2}\sum_{i,j=0}^{M-1} t_r^{i-j}
\end{equation}
Performing the integral over $\phi$, one then obtains:
\begin{equation}
\int d\phi P_r(\phi)\log(1-t_r)= \frac{-1}{M^2}\sum_{k=1}^{\infty} \sum_{i,j=0}^{M-1} \int \frac{t_r^{k+i-j}}{k} d\phi =\frac{-2\pi}{M^2}\sum_{k=1}^{\infty} \sum_{i,j=0}^{M-1} \frac{\delta_{k+i-j}}{k}= \frac{-2\pi}{M^2}\sum_{k=1}^M \frac{M-k}{k}
\end{equation}
Using harmonic sums, the sum on the RHS can be written as:
\begin{equation}
\frac{-2\pi}{M^2}\sum_{k=1}^M \frac{M-k}{k}=\frac{-2\pi}{M}(\log(M)+\gamma)+\frac{2\pi}{M} +o(\frac{1}{M^2})
\end{equation}
Summing over $r$ and noting that the two terms in the last line are identical after the sum and integral, we obtain, 
\begin{equation}
-4\pi \log(M)+4\pi (1-\gamma)
\end{equation}
Thus, the mutual information is,
\begin{equation}\label{mut info kit}
\boxed{
I=\log(M)-2(1-\gamma)}
\end{equation}

\subsubsection{RMSE} Let us consider the case where $\phi= \pi - \frac{\pi}{M}$, since that is where RMSE attains its maxima. The probabilities in \eqref{prob kit} are:
\begin{equation}
P_r = \frac{1}{M^2}\sec^2\left( \frac{(2r-1)\pi}{2M}\right)
\end{equation}
 Observing that $sec(\pi-x)=-sec(x)\implies P_{M+1-r}=P_r$, the expected value is given by
 \begin{equation}
 \sum_{r=1}^{M}\frac{2\pi (r-1)}{M}\sec^2\left( \frac{(2r-1)\pi}{2M}\right)= \sum_{r=1}^{M/2}\frac{2\pi (M-1)}{M}\sec^2\left( \frac{(2r-1)\pi}{2M}\right)=\pi-\frac{\pi}{M}
 \end{equation}
 We now show that the RMSE is $\sigma \sim \frac{1}{\sqrt{M}}$, to the leading order by evaluating it. The square of the RMSE, by definition, is given by
 \begin{equation}
 \sigma^2= \sum_{r=1}^{M}P_r \left(\frac{2\pi}{M} \right)^2 \left( \frac{M+1}{2}-r\right)^2
 \end{equation}
 The mean is near $r=\frac{M+1}{2}$. Therefore, changing the variable to $\frac{M+1}{2}-r$, we obtain a sum starting at the peak of the distribution:
 \begin{equation}\label{sigma_kit}
 \sigma^2= \frac{8\pi^2}{M^4}\sum_{r=1}^{\frac{M-1}{2}}r^2 \csc^2\left(\frac{r\pi}{M}\right)
 \end{equation}
The leading order in this sum is easily evaluated using an integral:
\begin{equation}
\frac{8\pi^2}{M^4}\sum_{r=1}^{\frac{M-1}{2}}r^2 \csc^2\left(\frac{r\pi}{M}\right)\approx  \frac{8}{M\pi}\int_{0}^{\frac{\pi}{2}} x^2 \csc^2(x)dx = \frac{8\log(2)}{M}
\end{equation}
Therefore, we have shown that, 
 \begin{equation}
 \boxed{
 \sigma = \sqrt{\frac{8\log(2)}{M}} \approx \frac{2.35}{\sqrt{M}}
 }
 \end{equation}
 
In general, the distance $|\phi-\frac{2\pi r}{M}|$ is incompatible with the topology of angles. A better choice is the length of the chord that subtends an angle $|\phi-\frac{2\pi r}{M}|$, given by $\sin^2(\frac{1}{2}(\phi-\frac{2\pi r}{M}))$. Using this distance, we obtain a similar order of magnitude for the RMSE, with the choice $\phi =  \pi + \frac{\pi}{M}$:

\begin{equation}
\sigma^2=\frac{1}{M^2}\sum_{r=0}^{M-1} \cos^2\left(\frac{(2r-1)\pi}{2M}\right)\sec^2\left( \frac{(2r-1)\pi}{2M}\right) = \frac{1}{M}
\end{equation}

\subsection{Q-Metrology}
In this strategy, a batch of $\nu$ identical repetitions is used to estimate every decimal place. Estimation with a precision of $1/N$ therefore requires $\log(N)$ batches of $\nu$ particles each. The precision capacity is $M=N^{\log(2\nu+1)}$
\subsubsection{Mutual Information}
The estimation of each decimal place is independent.  Therefore, since mutual information is additive, the total mutual information is the sum of the mutual informations evaluated for each batch. The distribution within a batch is binomial and therefore, we have:
\begin{equation}
\boxed{
I = \log(N)\left(\frac{1}{2}\log(\nu)\right) \approx \frac{1}{2}\log(M)
}
\end{equation} 
\subsubsection{RMSE}
The RMSE for the $i^{th}$ decimal place is $\frac{1}{\sqrt{\nu}}$. Therefore, the overall RMSE is given by
\begin{equation}
\boxed{
\sigma= \sqrt{\sum_{i=1}^{\log(N)}\frac{10^{-2i}}{\nu}} \approx \frac{1}{10\sqrt{\nu}}
}
\end{equation}

\end{document}